\documentstyle[12pt,amssymbols]{article}
\def\ltap{\raisebox{-.4ex}{\rlap{$\sim$}} \raisebox{.4ex}{$<$}}

\def\journal{\topmargin .3in	\oddsidemargin .5in
	\headheight 0pt	\headsep 0pt
	\textwidth 5.625in % 1.2 preprint size  %6.5in	
\textheight 8.25in % 1.2 preprint size 9in
	\marginparwidth 1.5in
	\parindent 2em
	\parskip .5ex plus .1ex		\jot = 1.5ex}
\journal

\newskip\humongous \humongous=0pt plus 1000pt minus 1000pt

\newif\ifdtup

\begin{document}
\begin{titlepage}
\begin{center}
\today \hfill    LBL- 34849\\
%          \hfill    UCB-PTH-XX/XX \\

\vskip .5in

{\large \bf Complementarity of Resonant and Nonresonant Strong $WW$ 
Scattering at the LHC}\footnote
{This work was supported by the Director, Office of Energy 
Research, Office of High Energy and Nuclear Physics, Division of High 
Energy Physics of the U.S. Department of Energy under Contract 
DE-AC03-76SF00098.}
%alternate footnote for faculty:
%\footnote{This work was supported in part by the Director, Office of 
%Energy Research, Office of High Energy and Nuclear Physics, Division of 
%High Energy Physics of the U.S. Department of Energy under Contract 
%DE-AC03-76SF00098 and in part by the National Science Foundation under 
%grant PHY90-21139.}

\vskip .5in

Michael S. Chanowitz and William Kilgore \\[.5in]

{\em Theoretical Physics Group\\
     Lawrence Berkeley Laboratory\\
     University of California\\
     Berkeley, California 94720}
%affiliation for faculty:
%{\em  Department of Physics\\
%      University of California\\
%      and\\
%      Theoretical Physics Group\\
%      Physics Division\\
%      Lawrence Berkeley Laboratory\\
%      1 Cyclotron Road\\
%      Berkeley, California 94720}
\end{center}

\vskip .5in

\begin{abstract}
We exhibit a complementary relationship between resonant $WZ$ and nonresonant
$W^+W^+$ scattering in a chiral Lagrangian model of the electroweak symmetry
breaking sector with a dominant ``$\rho$'' meson. We
use the model to estimate the minimum luminosity for the LHC to ensure a 
``no-lose'' capability to observe the symmetry breaking sector.
\end{abstract}

\end{titlepage}
%THIS PAGE (PAGE ii) CONTAINS THE LBL DISCLAIMER
%TEXT SHOULD BEGIN ON NEXT PAGE (PAGE 1)
\renewcommand{\thepage}{\roman{page}}
\setcounter{page}{2}
\mbox{ }

\vskip 1in

\begin{center}
{\bf Disclaimer}
\end{center}

\vskip .2in

\begin{scriptsize}
\begin{quotation}
This document was prepared as an account of work sponsored by the United
States Government.  Neither the United States Government nor any agency
thereof, nor The Regents of the University of California, nor any of their
employees, makes any warranty, express or implied, or assumes any legal
liability or responsibility for the accuracy, completeness, or usefulness
of any information, apparatus, product, or process disclosed, or represents
that its use would not infringe privately owned rights.  Reference herein
to any specific commercial products process, or service by its trade name,
trademark, manufacturer, or otherwise, does not necessarily constitute or
imply its endorsement, recommendation, or favoring by the United States
Government or any agency thereof, or The Regents of the University of
California.  The views and opinions of authors expressed herein do not
necessarily state or reflect those of the United States Government or any
agency thereof of The Regents of the University of California and shall
not be used for advertising or product endorsement purposes.
\end{quotation}
\end{scriptsize}

\vskip 2in

\begin{center}
\begin{small}
{\it Lawrence Berkeley Laboratory is an equal opportunity employer.}
\end{small}
\end{center}

\newpage
\renewcommand{\thepage}{\arabic{page}}
\setcounter{page}{1}
%THIS IS PAGE 1 (INSERT TEXT OF REPORT HERE)
%starthere
\noindent {\it \underline {Introduction} }

Partial wave unitarity implies that new quanta from the electroweak
symmetry breaking sector have masses
at or below about $ 4\sqrt{\pi / \sqrt{2}G_F} \simeq 2$ 
TeV.$^{\cite {mcmkg}}$ Since the bound is only a
rough estimate, the lightest new quanta could 
be a few times heavier, in which case they would not be directly 
observable at the LHC with any imaginable luminosity. The most effective
signal of the symmetry breaking sector would then be 
strong like-charge $WW$ scattering, $W^+W^+ + W^-W^-$.$^{\cite
{mcmkg,mbmc,jbetal}}$ We will consider the complementary relationship between 
resonant $J=1$ scattering in the $WZ$ channel and nonresonant scattering in the 
like-charge $WW$ channel. Unless there is a 
Higgs boson-like resonance, the $W^+W^+$
and/or $WZ$ channels provide much better prospects for detecting strong
scattering signals than the $ZZ$ channel.$^{\cite{mb-mc zz}}$

To incorporate the chiral symmetric dynamics of the electroweak symmetry
breaking sector, we use a chiral Lagrangian model$^{\cite{chrho}}$ 
with a TeV scale $I=J=~1$ $\rho$ resonance. Incorporating $SU(2)_L \times U(1)
_Y$ gauge symmetry, the model is equivalent to
the BESS model$^{\cite{bess}}$ (with $b=0$),
though we do not share the interpretation of $\rho$ as a gauge boson, choosing
instead to regard the model just as an effective Lagrangian for vector meson 
dominated, strongly coupled dynamics.
Applied to QCD$^{\cite{qcdapp}}$ we find (see figure 1) that the model fits
both $\pi^+\pi^0$ and $\pi^+\pi^+$ scattering data$^{\cite {pidata1,pidata2}}$ 
very well, to surprisingly high energy. (There is no tuning of parameters; 
the only inputs to figure 1 are the standard values of
$F_{\pi}$, $m_{\rho}$, and $\Gamma_{\rho}$.)

Applied to the electroweak sector with $m_{\rho}\ \ltap \ 2$ TeV,  the model 
implies a large resonant $WZ$ cross section and a
somewhat  suppressed $W^+W^+$ cross
section. For larger values of
$m_{\rho}$ the resonant $WZ$ signal decreases but nonresonant $W^+W^+$
scattering increases. This complementary relationship 
occurs because the chiral symmetry preserving contact interaction 
associated with cross-channel $\rho$ exchange 
suppresses $W^+W^+$ scattering, with less suppression for larger $m_{\rho}$.

In this paper we estimate the ``no-lose'' luminosity needed 
to provide an observable signal for {\em any} value of
$m_{\rho}$ in at least one of the two channels. We define a robust 
criterion for a
significant signal and then compute the minimum luminosity required to meet it.
We consider collider energies of 14 and 10 TeV, corresponding to the 
LHC design energy (with 1.8$^{\rm o}$ K magnets) in the first case and, 
in the last case, to the energy 
achievable in the LEP tunnel using existing (4.2$^{\rm o}$ K) magnet technology.
For the benefit of future generations (of archaeologists if not physicists) we
also present results for 40 TeV.

We find that the required luminosities are 60 and 190
fb$^{-1}$ for the LHC with 14 and 10 TeV respectively and 5 fb$^{-1}$ for 
40 TeV. These numbers codify the trade-off
between energy and luminosity for this class of physics, 
reflecting the energy dependence of both signal
and background cross sections. They do not reflect 
real-world complications that could effect the viability of high luminosity 
running such as event pile-up, instrumental radiation effects, or 
neutron-induced backgrounds.

A preliminary account of this work was presented 
previously.$^{\cite{mcdallas}}$. Similar results have been obtained 
by Bagger {\it et al.}$^{\cite{jbetal}}$. 
(One of several  important differences, 
reflected in the quoted signals, is that we have
optimized the cuts specifically for the chiral Lagrangian model 
for each particular value of $m_{\rho}$, as will
be done {\it in vivo} experimentally, while the signals quoted in 
reference \cite{jbetal} for all models refer to a single set of cuts chosen to 
optimize the signal for the standard Higgs boson model with $m_H=1$ TeV.)
Strong $WZ$ scattering signals with less
complete background studies have been reported 
previously.$^{\cite{mcmkg,chivukula,casalb2}}$
A more detailed account of our results will be presented 
elsewhere.$^{\cite{mcwk2}}$

\noindent{\it \underline {The Model} }

The naive $\rho\pi\pi$ interaction breaks chiral symmetry and 
violates the $\pi\pi$ low energy theorems. 
The minimal chiral invariant $\rho\pi\pi$ 
interaction$^{\cite{chrho}}$  contains an additional four pion
contact interaction that preserves the low energy theorems by cancelling 
the $\rho$ exchange contribution at threshold.
The partial wave amplitudes $a_{IJ}$ 
are then 
\begin{eqnarray}
a_{20} & = & {-\beta \over 32\pi}  \left\{ {s-2m_{\pi}^2 \over F_{\pi}^2}
        - {f_{\rho\pi\pi}^2 \over m_{\rho}^2} (2m_{\rho}^2 +3s -4m_{\pi}^2) 
            \right. \nonumber \\
       &   & \left.\hbox{     }\; \; \; + 2{f_{\rho\pi\pi}^2 \over \beta^2 s} 
               (m_{\rho}^2 + 2s -4m_{\pi}^2)
               {\rm ln}\left(1 + {\beta^2 s \over m_{\rho}^2}\right) \right\} \\
a_{11} & = & {\beta^3 \over 96\pi} \left\{ {s \over F_{\pi}^2} 
             -s{f_{\rho\pi\pi}^2 \over m_{\rho}^2}
             \left( {m_{\rho}^2 - 3s \over m_{\rho}^2 - s}\right) 
             \right. \nonumber \\
       &   & \left. \hbox{     }\; \; \; + 3{f_{\rho\pi\pi}^2 \over \beta^2 s}
             (m_{\rho}^2 + 2s -4m_{\pi}^2)\left[{-4 \over \beta^2}
             +{4m_{\rho}^2 + 2\beta^2 s\over \beta^4 s} {\rm ln}\left( 1 +
             {\beta^2 s \over m_{\rho}^2}\right) \right] \right\}
\end{eqnarray}
where $\beta$ is the pion velocity in the center of mass, $F_{\pi}=93$ MeV, and 
$f_{\rho\pi\pi}$ is determined from the $\rho$ width,
\begin{equation}
\Gamma_{\rho}= {f_{\rho\pi\pi}^2 \over 48\pi}\beta^3 m_{\rho},
\end{equation}
measured to be 151 MeV.

We unitarize these amplitudes by the K-matrix prescription
\begin{equation}
a_{IJ}^K = {\hbox{Re}(a_{IJ}) \over 1 - i\hbox{Re}(a_{IJ})}.
\end{equation}
For a resonant amplitude Re$(a)$ is evaluated with $\Gamma=0$, and 
$a^K$ is then equivalent to the
commonly used broad resonance prescription$^{\cite{pdg}}$ in which 
$\Gamma$ appears in the Breit-Wigner denominator evaluated at the center 
of mass energy, $\Gamma = \Gamma(s)$, rather than at the peak of the resonance. 
The resulting phase shifts, $a = e^{i\delta}\hbox{sin}\delta$, 
are compared with $\pi\pi$ data in figure 1. 
The agreement is very good, though it should not be taken
seriously above the resonance because of the {\it ad hoc} unitarization 
prescription and because 
the two contact interactions are really only known near threshold. 
The model also lacks an important element of $\pi\pi$ dynamics since it 
does not reproduce the
broad enhancement in the $a_{00}$ data (not shown) below 1 GeV. 

An important qualitative success of the model is its ability to reproduce 
the way in which the $I=2$ amplitude levels off above threshold. 
At threshold the $\rho$-induced contact interaction cancels the 
$\rho$ exchange contribution, leaving just the leading
chiral Lagrangian contact interaction that gives the low energy theorem. Away
from threshold in the $I=2$ channel the $\rho$-induced contact term grows 
faster than the $\rho$ exchange term;
it interferes destructively with the low energy theorem amplitude, causing
the $I=2$ amplitude to level off above threshold as observed in the data.

We are interested in the model not as a fully realistic representation 
of pion interactions, which despite figure 1 it surely is not, 
but as a tool to explore the relationship between
resonant and nonresonant strong $WW$ scattering.
We apply the model to the electroweak sector by replacing
$F_{\pi}$ with $v=246$ GeV and taking the Goldstone boson limit, 
$m_{\pi}=0$.\footnote{$W$ boson mass corrections are of the order of 
the corrections to the equivalence theorem. They are controlled by not
applying the model too close to the $WW$ threshold: the cuts used below ensure
that most of the signal is at $\sqrt{s_{WW}} > 500$ GeV. On the other hand, the
rapid decrease of the $WW$ effective luminosity as $s_{WW}$ increases ensures 
that the signals at the LHC are not dominated by $s_{WW}$ much larger than 
the domain of validity of the effective Lagrangian.}
The model is then completely specified by choosing $m_{\rho}$ and
$\Gamma_{\rho}$.

The range of possibilities is suggested by three cases. We consider minimal (one
doublet) $SU(4)$ and $SU(2)$ technicolor; using large $N_{TC}$ lore they imply
respectively $m_{\rho},\Gamma_{\rho} = 1.78,0.33$  and 2.52,0.92 in TeV. The
latter is the heaviest $\rho$ in conventional technicolor, since $m_{\rho}$
decreases as $N_{TC}$ and/or the number of techni-doublets are increased. To
present an even more difficult target we consider for our third case $m_{\rho},
\Gamma_{\rho} = 4.0,0.98$ TeV. The mass is set arbitrarily beyond the 
range of direct observability, and the width is fixed by taking
$f_{\rho\pi\pi}$ from the $\rho(770)$ of hadron physics. 

The $I,J=1,1$ and 2,0 partial waves are shown in figure 2. The
1.78 TeV $\rho$ provides the largest $I=1$ signal and the smallest 
for $I=2$, while the 4 TeV $\rho$ provides the smallest $I=1$ signal and 
the largest for $I=2$. As $m_{\rho}$ increases, both amplitudes
approach the K-matrix unitarization of the low energy
theorem amplitude (solid lines), 
from above for $I=1$ and from below for $I=2$.
If unobservable in one channel the signal may be observable in the other.

This complementarity follows from the suppression of the $I=2$ amplitude by the 
contact interaction associated with $\rho$ exchange as discussed above.
If $t$ and $u$ channel dynamics 
enhanced rather than suppressed the $I=2$ amplitude, the nonresonant limit would
be approached from above rather than from below as occurs here. The $W^+W^+$ 
signals 
would then be larger than they are in the model considered here, which in this
sense is a conservative model of the $I=2$ amplitude. 
In either case the amplitudes approach strong nonresonant
$WW$ scattering as $m_{\rho} \to \infty$, which becomes 
the signal of last resort if all quanta from the 
symmetry breaking sector are too heavy to produce directly.

\noindent{\it \underline {Signals:}}

Our criterion for a significant signal is 
\begin{equation}
\sigma^{\uparrow}   =  S/\sqrt{B}  \ge  5 
\end{equation}
\begin{equation}
\sigma^{\downarrow}   =  S/\sqrt{S+B}  \ge  3 
\end{equation}
\begin{equation}
S \ge B,
\end{equation}
where $S$ and $B$ are the number of signal and background events, and
$\sigma^{\uparrow}$ and $\sigma^{\downarrow}$ are respectively the number of 
standard deviations for the background to fluctuate
up to give a false signal or for the signal plus background to
fluctuate down to the level of the background alone. We apply these criteria 
below {\em after} the experimental acceptance is applied. 
In addition we require $S \ge B$ so that the signal is unambiguous despite the
systematic uncertainty in the size of the backgrounds, 
expected to be known to within $ \leq \pm 30 \%$ after ``calibration'' studies 
at the LHC.

For $WZ$ scattering we detect $WZ \rightarrow l\nu + \overline ll$
where $l = e$ or $\mu$, with net branching ratio $BR = 0.0143$.
The production mechanisms are $\overline qq \rightarrow \rho$, where the
$\rho\overline qq$ coupling has its origin in $W$--$\rho$ mixing\footnote
{Slightly different results follow from the 
$\rho$ dominance approximation, to be discussed in detail elsewhere.
$^{\cite{mcwk2}}$ } 
computed in the $SU(2)_L \times U(1)_Y$ gauged chiral Lagrangian 
${\cal L}_{\rm EFF}$$^{\cite{bess}}$, 
and $WZ$ fusion computed using the equivalence theorem$^{\cite{et1,mcmkg,et2}}$ 
and the effective $W$ approximation$^{\cite{ewa}}$ with
$a_{11}$ and $a_{20}$ unitarized as  described above. ($WZ$ scattering has a
resonant contribution from $a_{11}$ and a nonresonant contribution from
$a_{20}$.) The backgrounds are $\overline qq \rightarrow WZ$ and the complete 
$\hbox{O}(\alpha_W^2)$ amplitude for $qq \to qqWZ$. The latter is 
essentially the $qq \to qqWZ$ cross section from $SU(2) \times U(1)$ gauge 
interactions, computed in the
standard model with a light Higgs boson, say $m_H \le 0.1$ TeV. 

Requiring central lepton rapidity is both convenient experimentally and
helps to enhance the signal relative to the background: 
we require lepton rapidity $y_{l} <  2$. Cuts on the $Z$ transverse 
momentum, $p_{TZ} > p_{TZ}^{MIN}$,
and on the azimuthal angles $\phi_{ll}$ between the leptons from the $Z$ 
and the charged lepton from the $W$, $\hbox{cos}\phi_{ll} <
(\hbox{cos}\phi_{ll})^{MAX}$, are optimized for each choice of $m_{\rho}$ and
for each collider energy. We also examined the effect of a central jet
veto $^{\cite {vbetal}}$ on events with one or more hadronic jets 
with rapidity $y_j < 3$ and transverse momentum $p_{Tj} > 60$ GeV.\footnote
{For $WZ$ and $W^+W^+$ scattering the 
signal efficiency for the CJV was computed by taking the $m_H \to \infty$
limit of the standard model and imposing unitarity as in the linear model
of reference \cite {mcmkg}.}
Though it is included in the results quoted below, the CJV is not very 
effective against the $WZ$ backgrounds considered here since it does not reduce
the $\overline qq$ annihilation component of the background. It is likely to 
be useful, along with a lepton isolation requirement, against 
$t$ quark induced backgrounds that are not considered here but are shown
to be controllable by Bagger {\it et al}.$^{\cite{jbetal}}$

The detector efficiency for $WZ \rightarrow l\nu + \overline ll$ 
is estimated$^{\cite{sdctdr}}$ to be $0.85 \times 0.95 \simeq 0.8$.
Instead of correcting the theoretical cross sections, we take 
the acceptance into account 
by rescaling the significance criterion, replacing equations (5) and
(6) by $\sigma^{\uparrow} \ge 5.5$ and $\sigma^{\downarrow} 
\ge 3.3$. Table 1 displays ${\cal L}_{MIN}$, the minimum luminosity needed to
satisfy the criterion, for each model and collider energy. Also displayed 
are the signal and background cross sections and the optimal values of 
$p_{TZ}^{MIN}$ and $(\hbox{cos}\phi_{ll})^{MAX}$. 
For $m_{\rho} = 4$ TeV  no signal is indicated for the 
LHC with either 14 or 10 TeV, because there are no cuts that 
satisfy $S\ge B$. 

The $W^+W^+$ channel has the largest leptonic branching ratio,
$\simeq 0.05$ to $e$'s and/or $\mu$'s, and no ${\overline q}q$ 
annihilation background.
The signature is striking: two isolated, high $p_T$, like-sign leptons in an
event with no other significant activity (jet or lepton) in the central region.
The dominant backgrounds are the $\hbox{O}(\alpha_W^2)$$^{\cite{dv2}}$ and 
$\hbox{O}(\alpha_W
\alpha_S)$$^{\cite{mcmg-dv1}}$ amplitudes for $qq \rightarrow qqWW$. 
The former, like the analogous $WZ$ background discussed above, is 
computed in the standard model with a light Higgs boson.
Other backgrounds, from $W^+W^-$ with
lepton charge mismeasured and from $\overline tt$ production, require detector 
simulation. Studies in the SDC TDR$^{\cite{sdctdr}}$ show that they can be 
controlled, at least for ${\cal L}=10^{33}$ cm$^{-2}$ sec$^{-1}$ 
(see also reference \cite{t-w}).

A powerful set of cuts that indirectly exploits the 
longitudinal polarization of the signal has emerged from the efforts of three
collaborations.$^{\cite {mbmc,vbetal,dgv}}$. The most useful cuts are on the 
lepton transverse
momentum $p_{Tl}$, on the azimuthal angle between the two leptons 
$\phi_{ll}$, and a veto on events with central jets as defined above.
With just the lepton rapidity cut $y_l < 2$ the
background is an order of magnitude bigger than the largest of the signals;
the additional cuts typically reduce the background by factors of order 200 or
300 while decreasing the signal by factors of only 2 or  3. 

Assuming 85\% detection efficiency for a single isolated 
lepton,$^{\cite{sdctdr}}$ the significance criterion, inequalities 
(5--6), applied to the uncorrected yields become 
$\sigma^{\uparrow}>6$ and $\sigma^{\downarrow}>3.5$. The minimum luminosities
to meet this criterion are summarized in table 2. 

\noindent{\it \underline {Discussion}}

Table 1 shows that the 1.78 TeV $\rho$ would be observable in $WZ$ scattering 
at the LHC with 44
fb$^{-1}$ and could even be observed at a 10 TeV collider with 120 fb$^{-1}$.
The 4 TeV $\rho$ cannot
be distinguished from nonresonant strong scattering, and it offers no signal 
at the LHC in the $WZ$ channel satisfying inequality (7). 

Nonresonant scattering is more readily observed in the $W^+W^+ + W^-W^-$
channel; at the LHC the like-charge $W$ pair signal for $m_{\rho}=4$ TeV 
meets the criterion with 48
fb$^{-1}$. The smaller cross section for $m_{\rho}=1.78$ TeV would be
observable with 86 fb$^{-1}$. A 10 TeV collider would require 150 and 240
fb$^{-1}$ respectively. 

The worst case scenario is represented (roughly speaking\footnote{ 
A rough exploration of $m_{\rho},\Gamma_{\rho}$ parameter space reveals
cases somewhat worse but not dramatically different than the 2.52 TeV $\rho$
considered here.})
by the 2.52 TeV $\rho$ meson: it is
heavy enough to present a small resonant signal in the $WZ$ channel but light
enough to effectively suppress nonresonant scattering in the 
$W^+W^+$ channel. The best signal is in the $W^+W^+ + W^-W^-$ channel, 
where 63 fb$^{-1}$ provide 
a signal meeting our criterion. This defines the ``no-lose
luminosity'', since it ensures a significant signal for {\em any} value 
of $m_{\rho}$ in at least one of the two channels. 
For 10 and 40 TeV colliders the corresponding ``no-lose''
luminosities are 190 and 5 fb$^{-1}$ respectively.

We have not included top quark related backgrounds. They are distinguished
from the signals by higher jet multiplicities and by lepton isolation criteria.
Theoretical$^{\cite{jbetal}}$  and experimental$^{\cite{sdctdr}}$ 
simulations suggest they will not
dramatically alter the conclusions reported here, though the experimental 
simulations were for 10$^{33}$cm$^{-2}$sec$^{-1}$ luminosity and should be
reconsidered for higher luminosity.

The $ZZ$ channel
provides the best signal for scalar resonances such as a heavy Higgs boson,
but is less useful for vector resonances or nonresonant scattering.
Including the gluon-gluon
fusion component with $m_t = 150$ GeV,
the nonresonant strong scattering signal (for the ``linear
model''$^{\cite{mcmkg}}$) is only just 
observable with 10 fb$^{-1}$ at a 40 TeV collider and 
would require $\simeq 350$ fb$^{-1}$ at a 16 TeV 
collider.$^{\cite{mb-mc zz}}$\footnote{These 
results refer to the ``silver-plated'' channel, $ZZ \rightarrow
\overline ll + \overline \nu \nu$. } 

The results presented here for $W^+W^+$ and $WZ$ scattering are encouraging 
if luminosities of
order 10$^{34}$cm$^{-2}$sec$^{-1}$ can be achieved and if they can be 
used. Detector simulations, especially of the $W^+W^+$ channel, are  needed to
establish feasibility at the necessary luminosity.

\noindent Acknowledgements: Some of the computations were performed on a MASPAR
parallel processor, courtesy of the the LBL Information \& Computing Sciences
Division. This work was supported 
by the Director, Office of Energy 
Research, Office of High Energy and Nuclear Physics, Division of High 
Energy Physics of the U.S. Department of Energy under Contract 
DE-AC03-76SF00098.

\newpage
{\center \bf Tables \\}
\vskip .5in

%Table 1
\begin{small}
\begin{quotation}
Table 1. 
Minimum luminosity to satisfy observability criterion for $W^{\pm}Z$ scattering
for $\sqrt{s}=10,14,40$ TeV and $m_{\rho}=1.78, 2,52,4.0$
TeV. Each entry contains ${\cal L}_{MIN}$ in fb$^{-1}$, the number of
signal/background events per 10 fb$^{-1}$, and the corresponding values of
the cut parameters $p_{TZ}^{MIN},{\rm cos}(\phi_{ll})^{\rm MAX}$. A central jet
veto is applied as discussed in the text.
\end{quotation}
\end{small}

\begin{center}
\begin{tabular}{c|c|c|c}
   & 1.78 TeV & 2.52 TeV & 4.0 TeV \\
\hline   
    &    &   &   \\
    &  120 fb$^{-1}$& 1400 fb$^{-1}$ & No   \\
10 TeV  &1.4/0.71   & 0.15/0.11 & \\
    & 475 GeV,1.0  & 675 GeV,0.9 & Signal \\
    &    &   &   \\
\hline
    &    &   &   \\
    & 44 fb$^{-1}$  & 300 fb$^{-1}$ & No \\
14 TeV  &3.8/2.0   & 0.58/0.34 & \\
    & 450 GeV,1.0 & 675 GeV,1.0 & Signal \\
    &    &   &   \\
\hline
    &    &   &   \\
    &4.8 fb$^{-1}$  & 12 fb$^{-1}$ & 35 fb$^{-1}$ \\
40 TeV  &33/19 & 13/7.0 & 5.2/3.1 \\
    &400 GeV,1.0  & 500 GeV,$-0.4$ & 650 GeV,0.8 \\
    &    &   &   \\
\hline
\end{tabular}
\end{center}

\newpage
%Table 2
\begin{small}
\begin{quotation}
Table 2. 
Minimum luminosity to satisfy observability criterion for $W^+W^+ + W^-W^-$ 
scattering for $\sqrt{s}=10,14,40$ TeV and $m_{\rho}=1.78$, 2,52, 4.0 
TeV. Each entry contains ${\cal L}_{MIN}$ in fb$^{-1}$, the number of
signal/background events per 10 fb$^{-1}$, and the corresponding values of
the cut parameters $p_{Tl}^{MIN},{\rm cos}(\phi_{ll})^{\rm MAX}$. A central jet
veto is applied as discussed in the text.
\end{quotation}
\end{small}

\begin{center}
\begin{tabular}{c|c|c|c}
   & 1.78 TeV & 2.52 TeV & 4.0 TeV \\
\hline   
    &    &   &   \\
    & 240 fb$^{-1}$ & 190 fb$^{-1}$ & 150 fb$^{-1}$ \\
10 TeV  & 0.79/0.42 & 1.0/0.53 & 1.2/0.59 \\
    & 90 GeV,$-0.8$ & 80 GeV,$-0.85$ & 80 GeV,$-0.80$ \\
    &    &   &   \\
\hline
    &    &   &   \\
    & 86 fb$^{-1}$ & 63 fb$^{-1}$ & 48 fb$^{-1}$ \\
14 TeV  & 2.2/1.2 & 2.9/1.5 & 3.9/2.0 \\
    & 80 GeV,$-0.875$ & 70 GeV,$-0.875$ & 70 GeV,$-0.725$ \\
    &    &   &   \\
\hline
    &    &   &   \\
    & 7.4 fb$^{-1}$ & 5.2 fb$^{-1}$ & 3.6 fb$^{-1}$ \\
40 TeV  & 25/12 & 33/12 & 44/12 \\
    & 80 GeV,$-0.75$ & 80 GeV,$-0.75$ & 80 GeV,$-0.75$ \\
    &    &   &   \\
\hline

\end{tabular}
\end{center}

\newpage
\begin{center}
{\bf Figure Captions}
\end{center}
\vskip .5 in
\noindent Figure 1. The effective Lagrangian model, ${\cal L}_{EFF}$,
compared with $\pi\pi$
scattering data$^{\cite{pidata1,pidata2}}$ for $|a_{11}|$ and $\delta_{20}$.
\vskip .3 in
\noindent Figure 2. $|a_{11}|$ and $|a_{20}|$ for the effective Lagrangian
applied to the electroweak symmetry breaking sector 
with $m_\rho = 1.78$ (dashes), $m_\rho = 2.52$ (long dashes) and
$m_\rho = 4.0$ TeV (dot-dash).  The nonresonant $K$-LET model is indicated 
by the solid lines.

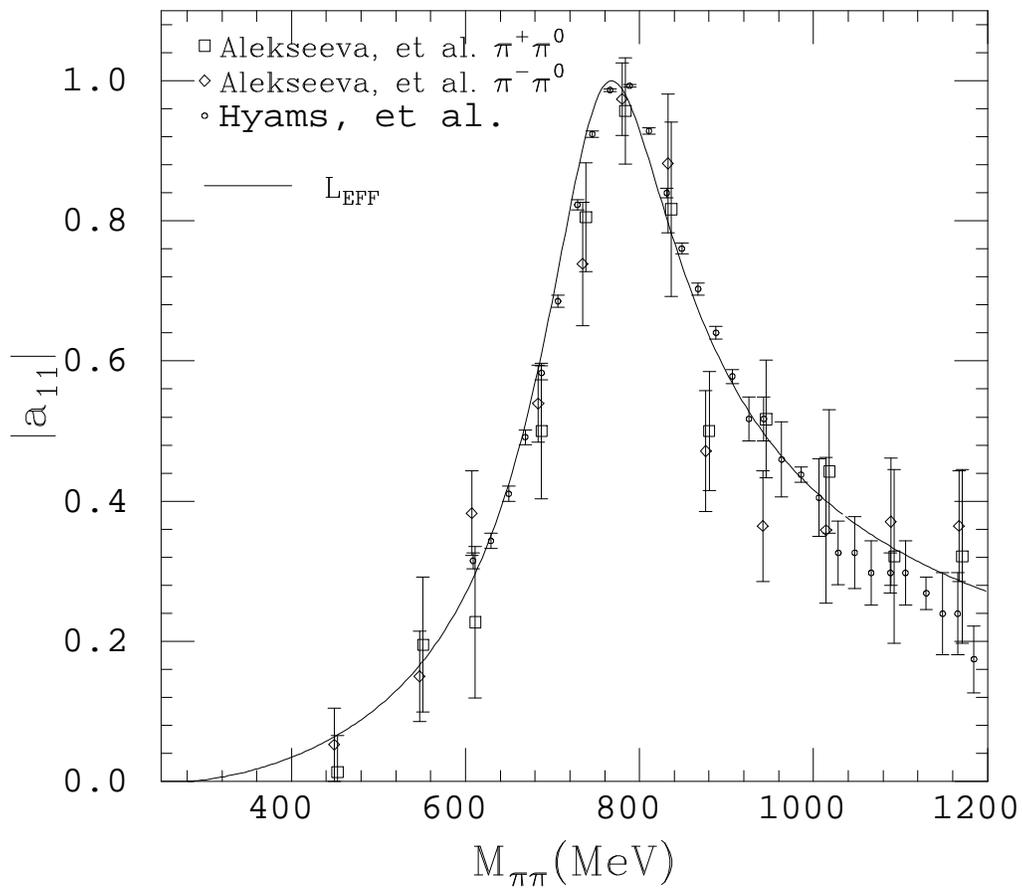
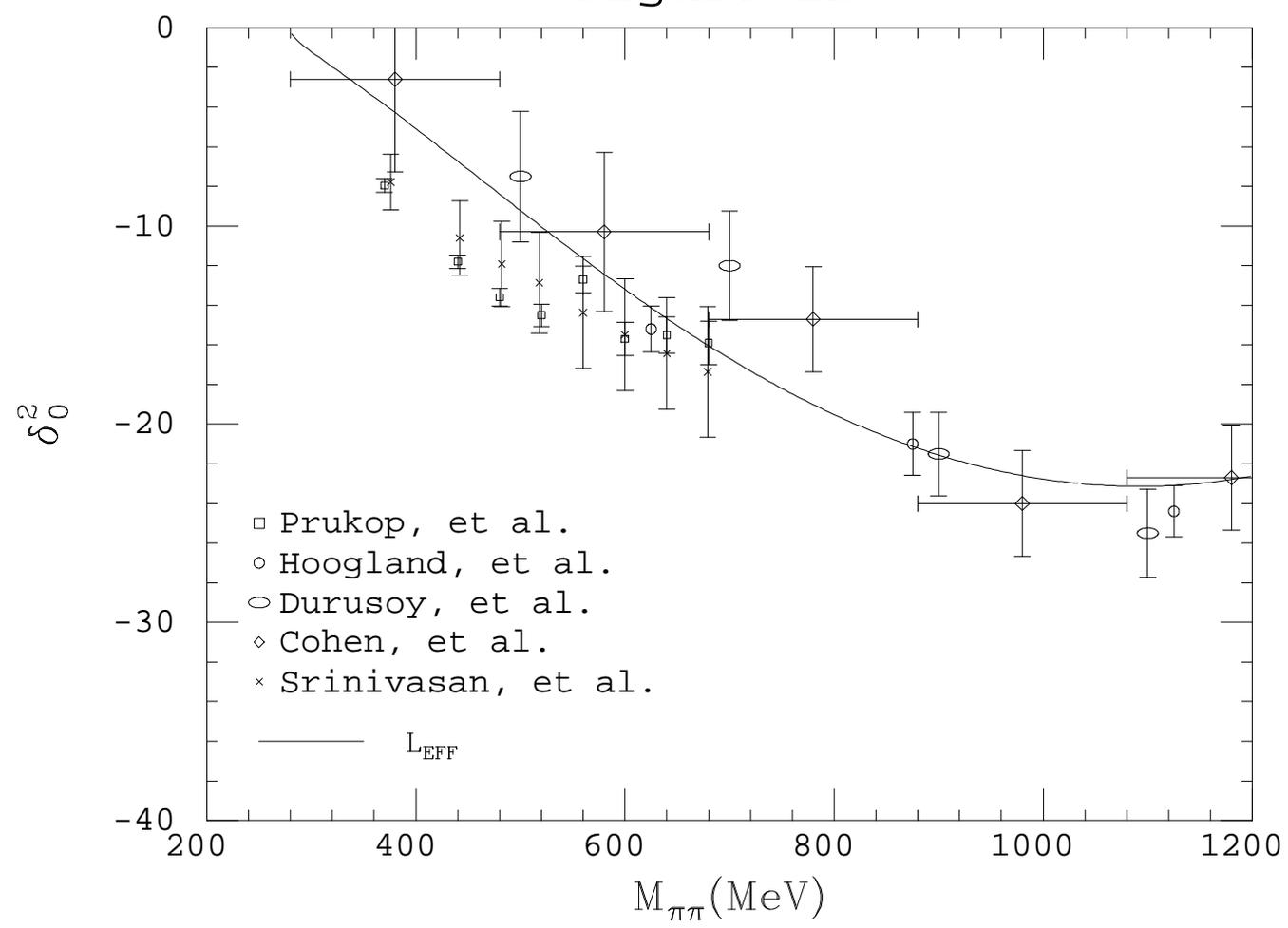
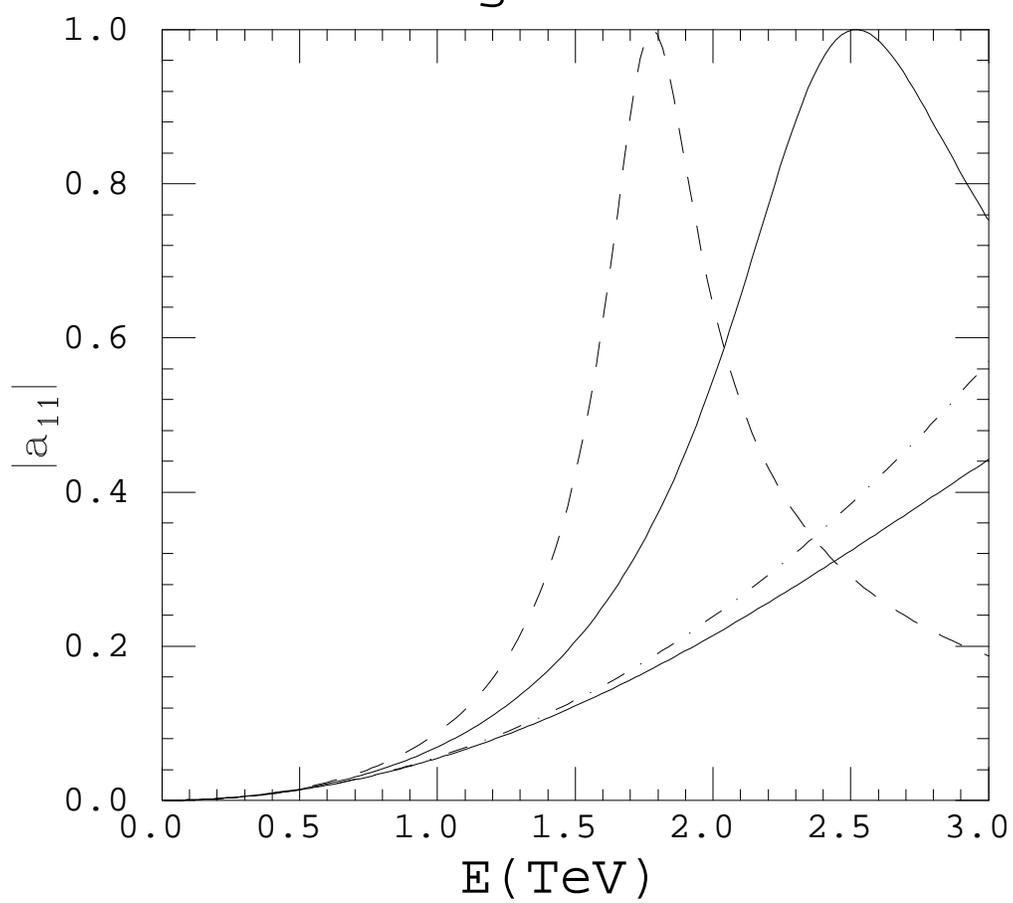
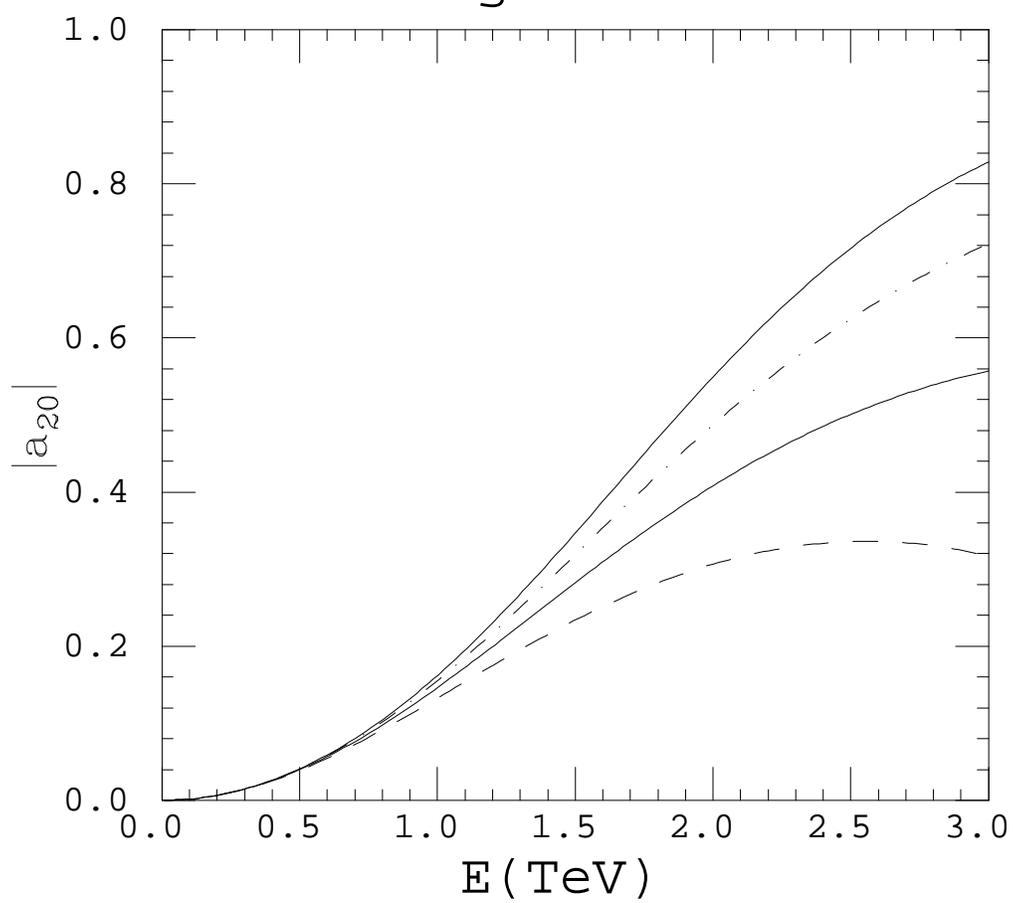
\end{document}